\documentstyle[aps,prl,times]{revtex}
\newcommand{\pitoee}{\mbox{$\pi^0 \rightarrow e^+e^- $}}
\newcommand{\pidal}{\mbox{$\pi^0 \rightarrow e^+e^-\gamma $}}
\newcommand{\piddal}{\mbox{$\pi^0 \rightarrow e^+e^-e^+e^- $}}
\newcommand{\kthreepz}{\mbox{$K_L \rightarrow 3 \pi^0 $}}
\newcommand{\kwx}{\mbox{$(m_{e^+e^-}/m_{\pi^0})^2$}}

\newcommand{\mee}{\mbox{$m_{e^+e^-}$}}
\newcommand{\meeg}{\mbox{$m_{e^+e^-\gamma}$}}
\newcommand{\mpipiee}{\mbox{$m_{\pi^0\pi^0 e^+e^-}$}}

\newcommand{\pigsgs}{\mbox{$\pi^0 \gamma^{\ast}\gamma^{\ast}$}}
\newcommand{\pigg}{\mbox{$\pi^0 \rightarrow \gamma\gamma $}}

\input{psfig}
\begin{document}
\title{Measurement of the branching ratio of \pitoee\ using \kthreepz\
decays in flight}
\draft
%\address{University of Batavia, Batavia, Illinois, 65432}
\author{
A.~Alavi-Harati$^{12}$,
I.F.~Albuquerque$^{10}$,
T.~Alexopoulos$^{12}$,
M.~Arenton$^{11}$,
K.~Arisaka$^2$,
S.~Averitte$^{10}$,
A.R.~Barker$^5$,
L.~Bellantoni$^7$,
A.~Bellavance$^9$,
J.~Belz$^{10}$,
R.~Ben-David$^7$,
D.R.~Bergman$^{10}$,
E.~Blucher$^4$, 
G.J.~Bock$^7$,
C.~Bown$^4$, 
S.~Bright$^4$,
E.~Cheu$^1$,
S.~Childress$^7$,
R.~Coleman$^7$,
M.D.~Corcoran$^9$,
G.~Corti$^{11}$, 
B.~Cox$^{11}$,
M.B.~Crisler$^7$,
A.R.~Erwin$^{12}$,
R.~Ford$^7$,
A.~Golossanov$^{11}$,
G.~Graham$^4$, 
J.~Graham$^4$,
K.~Hagan$^{11}$,
E.~Halkiadakis$^{10}$,
K.~Hanagaki$^8$,  
S.~Hidaka$^8$,
Y.B.~Hsiung$^7$,
V.~Jejer$^{11}$,
J.~Jennings$^2$,
D.A.~Jensen$^7$,
R.~Kessler$^4$,
H.G.E.~Kobrak$^{3}$,
J.~LaDue$^5$,
A.~Lath$^{10}$,
A.~Ledovskoy$^{11}$,
P.L.~McBride$^7$,
A.P.~McManus$^{11}$,
P.~Mikelsons$^5$,
E.~Monnier$^{4},$%\footnote[0]{$^*$ On leave from C.P.P. Marseille/C.N.R.S., France},
T.~Nakaya$^7$,
U.~Nauenberg$^5$,
K.S.~Nelson$^{11}$,
H.~Nguyen$^7$,
V.~O'Dell$^7$, 
M.~Pang$^7$, 
R.~Pordes$^7$,
V.~Prasad$^4$, 
C.~Qiao$^4$, 
B.~Quinn$^4$,
E.J.~Ramberg$^7$, 
R.E.~Ray$^7$,
A.~Roodman$^4$, 
M.~Sadamoto$^8$, 
S.~Schnetzer$^{10}$,
K.~Senyo$^8$, 
P.~Shanahan$^7$,
P.S.~Shawhan$^4$, 
W.~Slater$^2$,
N.~Solomey$^4$,
S.V.~Somalwar$^{10}$, 
R.L.~Stone$^{10}$, 
I.~Suzuki$^8$,
E.C.~Swallow$^{4,6}$,
R.A.~Swanson$^{3}$,
S.A.~Taegar$^1$,
R.J.~Tesarek$^{10}$, 
G.B.~Thomson$^{10}$,
P.A.~Toale$^5$,
A.~Tripathi$^2$,
R.~Tschirhart$^7$, 
Y.W.~Wah$^4$,
J.~Wang$^1$,
H.B.~White$^7$, 
J.~Whitmore$^7$,
B.~Winstein$^4$, 
R.~Winston$^4$, 
J.-Y.~Wu$^5$,
T.~Yamanaka$^8$,
E.D.~Zimmerman$^{4,*,\dagger}$\footnote[0]{$^{*}$ To whom correspondence should be addressed. Electronic address: edz@fnal.gov.}\footnote[0]{$^\dagger$ Current address: Columbia University, New York, New York 10027.}
}
%\vspace*{0.1in}
%\footnotesize
\address{
$^1$ University of Arizona, Tucson, Arizona 85721 \\
$^2$ University of California at Los Angeles, Los Angeles, California 90095 \\
$^{3}$ University of California at San Diego, La Jolla, California 92093 \\
$^4$ The Enrico Fermi Institute, The University of Chicago, 
Chicago, Illinois 60637 \\
$^5$ University of Colorado, Boulder, Colorado 80309 \\
$^6$ Elmhurst College, Elmhurst, Illinois 60126 \\
$^7$ Fermi National Accelerator Laboratory, Batavia, Illinois 60510 \\
$^8$ Osaka University, Toyonaka, Osaka 560 Japan \\
$^9$ Rice University, Houston, Texas 77005 \\
$^{10}$ Rutgers University, Piscataway, New Jersey 08855 \\
$^{11}$ The Department of Physics and Institute of Nuclear and 
Particle Physics, University of Virginia, 
Charlottesville, Virginia 22901 \\
$^{12}$ University of Wisconsin, Madison, Wisconsin 53706 \\
%{\rm (E799-II Collaboration)\\}
}
\normalsize

\date{\today}

\maketitle

%\vspace{0.1in}
%{\center\Large COLLABORATION DRAFT, \today \\}
%\vspace{0.1in}

\begin{abstract}
The branching ratio of the rare decay \pitoee\ has been measured
in E799-II, a rare kaon decay experiment using the KTeV detector at 
Fermilab. %The $\pi^0$'s were produced in fully-reconstructed \kthreepz\ 
%decays in flight. 
We observed 275 candidate 
\pitoee\ events, with an expected background of $21.4\pm 6.2$ events 
which includes the contribution from Dalitz decays. We measured
$
{\rm BR}\left (\pitoee , \kwx >0.95 \right) =
           (6.09 \pm 0.40 \pm 0.24) \times 10^{-8},
$
where the first error is statistical and the second systematic. This
result is the first significant observation of the excess rate for this decay
above the unitarity lower bound. %It is in agreement with expectations 
%from \pigsgs\ form factor models and recent chiral perturbation theory 
%calculations.

\vspace{0.1in}

\noindent
PACS numbers: 13.20.Cz, 13.40.Gp, 13.40.Hq

\end{abstract}

\vspace{0.1in}

\newpage
\narrowtext
\twocolumn
%\linespace{2.0}

The decay \pitoee\ has received much experimental and theoretical attention
since its branching ratio was first calculated by Drell in
1959 \cite{drell}. Within the Standard Model, this decay proceeds
predominantly through a two-photon intermediate state, at
a rate less than $10^{-7}$ that of $\pi^0 \rightarrow \gamma\gamma$.
Relative to \pigg , \pitoee\ is suppressed by a helicity factor 
$(2m_e/m_{\pi^0})^2$ as well as by two orders of $\alpha_{\rm EM}$. 
The contribution
to the decay from on-shell internal photons has been calculated exactly
in QED \cite{berman}, and forms the ``unitarity bound,'' a lower limit
on the branching ratio which is 
${\rm BR}(\pitoee )\geq 4.69\times 10^{-8}$, ignoring final-state radiative
effects. The contribution from off-shell photons depends on the
\pigsgs\ form factor, and is model-dependent. Recent 
vector meson dominance \cite{ametller} and chiral perturbation theory 
\cite{savage,pich} calculations predict 
this contribution to be somewhat smaller than that from on-shell photons, 
giving a total branching ratio of $(6-9)\times 10^{-8}$, ignoring 
radiative corrections. 

     Earlier experiments have produced conflicting measurements of the 
     branching ratio for this mode. The earliest measurements of \pitoee\ 
     were performed by a Geneva-Saclay group in 1978 \cite{genevasaclay}
     using $\pi^0$'s produced by the decay $K^+ \rightarrow \pi^+\pi^0$ 
     in flight, and by a LAMPF group in 1983 \cite{lampfee} using the 
     charge-exchange process $\pi^-p \rightarrow \pi^0n$ from a 300~MeV/$c$
     pion beam. Both experiments favored a branching ratio of 
     $\sim 2\times 10^{-7}$, which would be hard to accomodate within 
     the Standard Model. A 1989 search by the SINDRUM collaboration 
     \cite{sindrum}, using stopped $\pi^-p \rightarrow \pi^0n$, produced 
     a 90\% confidence level upper limit of $1.3\times 10^{-7}$, excluding 
     the central values of both previous measurements. In 1993, BNL 
     E851 \cite{bnl851} and FNAL E799-I \cite{ksmcfprl} observed
     the decay at the $(5-10)\times 10^{-8}$ level, near the Standard Model
     expectation. The BNL measurement used $K^+ \rightarrow \pi^+\pi^0$ 
     decays, while the FNAL experiment used \kthreepz\ decays.

In this Letter we present a new, precision measurement of 
${\rm BR}(\pitoee )$ from E799-II, a rare $K_L$ decay experiment which 
took data in 1997 %as part of the KTeV program 
at Fermilab. The $\pi^0$'s were produced using
\kthreepz\ decays in flight, where the other two $\pi^0$'s in the event
decayed to $\gamma\gamma$. The \pitoee\ events were normalized to 
Dalitz decays (\pidal ) with $\mee > 65$~MeV/$c^2$,
which were collected and analyzed simultaneously. High-\mee\ events 
were used in order to keep the charged track kinematic variables as similar as
possible for the signal and normalization modes, and thus cancel many
detector-related systematic errors.

This technique, which was adapted from E799-I, has significant 
advantages over those used in other 
measurements. The $\pi^0$'s were produced and decayed in vacuum, eliminating
backgrounds and resolution smearing
from decay products scattering or converting in charge-exchange
targets. The continuum process $K_L\rightarrow \pi^0\pi^0 e^+e^-$ has 
never been observed and does not pose a significant background. By contrast,
the analogous processes in $K^+$ experiments ($K^+ \rightarrow \pi^+e^+e^-$)
and charge-exchange experiments ($\pi^-p \rightarrow e^+e^-X$) both 
produce large backgrounds to \pitoee . Reconstruction of the full kaon
decay provides redundant kinematic constraints, eliminating all non-\kthreepz\ 
backgrounds.

\begin{figure}
  \centerline{ \psfig{figure=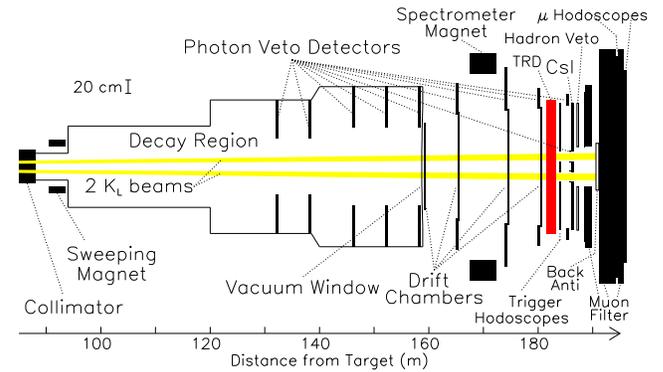,width=8.5cm}}
  \caption{Plan view of the KTeV spectrometer as configured for E799-II. 
           The horizontal scale is compressed.
\label{spectrometer}}
\end{figure}

The elements of the E799-II spectrometer (Fig.~\ref{spectrometer})
relevant to this measurement are described below. 
Two nearly-parallel neutral kaon beams were produced
by 800~GeV protons striking a 30~cm BeO target at a targeting angle of 
4.8~mrad. Two neutral beams, each one up to 0.35~$\mu$sr, were
defined by collimators. A Pb absorber converted photons in the beam, and
charged particles were removed by a series of sweeping magnets. A 65~m 
evacuated decay region ended at a Mylar-Kevlar vacuum
window 159~m from the target. The beams in the decay region were 
composed mostly
of neutrons and $K_L$, with small numbers of $K_S$, $\Lambda^0$,
$\bar \Lambda^0$, $\Xi^0$, $\bar \Xi^0$. These short-lived particles
tended to decay upstream. The $K_L$ momentum ranged from $\sim 20$ to
$\sim 200$~GeV/$c$. %; approximately 5\% of the $K_L$ decayed before reaching
%the end of the decay region. 

%Immediately downstream of the vacuum window was a charged particle
%spectrometer 
%consisting of four drift chambers with two orthogonal views. 
%A dipole magnet downstream of the second chamber provided a 
%transverse momentum kick of 205~MeV/$c$ in the horizontal direction. 
%Charged particle resolution in the chambers was
%$\sim 100$~$\mu$m. Helium bags were placed between the chambers to 
%reduce multiple scattering and photon conversions. 
       Charged particles were tracked using four drift chambers with two 
       orthogonal views; a dipole magnet downstream of the second chamber 
       provided a transverse momentum kick of 205~MeV/$c$. Helium bags were
       placed between the chambers to reduce multiple scattering and photon
       conversions.

Photon energy measurement and electron identification were performed 
using a 3100-block pure CsI electromagnetic calorimeter. % \cite{ajrcsi}. 
The photon energy resolution was $\sim 1$\%, averaged over the energy range 
typical of \pitoee\ events (2$-$60~GeV). Immediately upstream of the CsI,
two overlapping banks of
scintillation counters (the ``trigger hodoscopes'') provided fast signals
for triggering on charged particles. Downstream of the calorimeter, a
10~cm lead wall followed by a scintillator plane formed a hadron veto
which rejected at trigger level events with charged pions in the final state. 
An eleven-plane photon veto system, consisting of 
lead-scintillator counters throughout the decay region and spectrometer,
detected particles which left the fiducial volume.

%A single trigger collected both \pitoee\ candidates and \pidal\ 
%events for normalization. 
The trigger required at least 24~GeV of in-time 
energy in the CsI, and hits in the drift chambers and trigger hodoscopes 
consistent with at
least two tracks. Events were rejected when more than 0.5~GeV was
deposited in any photon veto counter, or more than the equivalent of
2.5 minimum ionizing particles were detected in the hadron veto. 
A hardware processor \cite{hcc} required at least four energy clusters
in the CsI, where a ``cluster'' was a set of contiguous blocks with
at least 1~GeV deposited in each block. 

Offline, events with exactly two reconstructed tracks were selected and the
tracks were required to form a common 
vertex inside the decay region. They also had to be electron candidates,
defined as tracks which pointed to a CsI cluster whose energy $E$ was 
within $\pm$8\% of the track momentum $p$. The reconstructed kaon energy 
was required to be at least 40~GeV, and each cluster in the CsI at least 
1.5~GeV. These cuts reduced the dependence of the result on 
CsI trigger thresholds. Clusters without tracks pointing 
to them were assumed to be photons. \pitoee\ candidates were
required to have four photons; \pidal\ candidates were required to have five.

Photons were reconstructed assuming that each pair of photons
came from a \pigg\ decay. We calculated the distance 
$Z_{12} \equiv (r_{12}/m_{\pi^0}) \sqrt{E_1 E_2}$ of the decay from
the CsI, where $E_i$ is the energy of photon $i$ and $r_{ij}$ is the 
transverse separation of photons $i$ and $j$ at the CsI. 
%%For each pairing
%%of the photons into two $\pi^0$'s, we calculated the ``pairing $\chi^2$'' 
%%for the hypothesis that the two $\pi^0$ decays occurred at the same 
%%position.
%%%$$
%%%\chi^2 = \frac{(Z_{12} - Z_{34})^2}{\delta_{Z_{12}}^2 + \delta_{Z_{34}}^2}
%%%$$
%%%where $\delta_{Z_{ij}}$ is the resolution of $Z_{ij}$. 
For %four-photon \kthreepz , 
\pitoee\ candidates, there were three possible pairings of the 
photons into two $\pi^0$'s. The pairing was chosen which minimized the 
$\chi^2$ for the hypothesis that the two $\pi^0$ decays occurred at the
same position. For %the five-photon \kthreepz , 
\pidal\ candidates, there were fifteen pairings; the best 
was selected and the unpaired photon was assumed to have come from 
the \pidal\ decay. 

%After selecting a photon pairing, a weighted average was taken of the 
%$Z$ positions of the two $\pi^0$'s. 
The photon four-momenta were calculated
assuming the photons originated at the weighted average of the $Z$ positions
of the two $\pi^0$'s and the transverse
position of the reconstructed two-electron vertex. 
This reconstruction method allowed
kinematic quantities to be calculated in nearly the same way for the
signal and normalization modes, thereby canceling certain systematic errors.

The basic reconstruction cuts described below were applied, and the samples
obtained were used to study acceptance and backgrounds. The total invariant
mass \mpipiee\ was required to be within 50~MeV/$c^2$ of the 
$K_L$ mass. The total momentum transverse to the kaon direction was required 
to be less than 
30~MeV/$c$. For the normalization sample, the Dalitz decay mass \meeg\ 
was required to be within 30~MeV/$c^2$ of the $\pi^0$
mass, and the reconstructed electron pair mass $\mee > 70$~MeV/$c^2$ 
in order to avoid systematic errors from mass resolution smearing near 
the 65~MeV/$c^2$ cutoff. 

A detailed Monte Carlo (MC) simulation was used to estimate acceptance
for the signal and normalization modes, as well as the level of
backgrounds in the samples. Both the signal and normalization MC were
implemented with radiative corrections.% based on analytic calculations. 
The \pitoee\ MC used the ${\cal O}(\alpha_{\rm EM})$ 
radiation model of Bergstr\"om \cite{bergstrom} , and the \pidal\ MC 
used an ${\cal O}(\alpha_{\rm EM}^2)$ calculation based on the work of 
Mikaelian and Smith \cite{mikaelian}.

\begin{figure}
  \centerline{ \psfig{figure=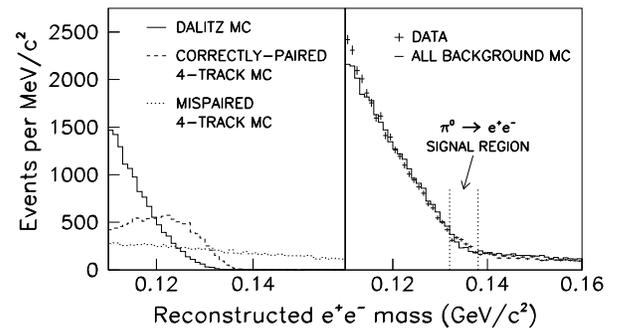,width=8.0cm}}
  \caption{Reconstructed \mee\ after basic reconstruction cuts. 
           Left: background MC predictions. Right: E799-II data 
           overlaid on sum of background predictions. (Backgrounds
           are normalized to the observed number of fully-reconstructed
           \pidal\ decays.)
\label{bkloose_prl}}
\end{figure}

At this stage, the sample in the \pitoee\ signal region 
($0.132 < \mee < 0.138$~GeV/c$^2$) was background-dominated.
Fig. \ref{bkloose_prl} shows the distribution of \mee\ for data
and for the MC background predictions. The backgrounds, which all 
came from \kthreepz\
decays, were as follows. Very high-\mee\ Dalitz decays (\pidal ) could
be misreconstructed as \pitoee\ if the photon was not detected and \mee\ was
reconstructed slightly high (by 1-10~MeV). Another
type of background came from decays with four electrons in the final
state. When one electron of each sign was soft, the spectrometer magnet 
could sweep them out of the fiducial volume, leaving only two reconstructible
tracks. The four electrons could come from \kthreepz\ with multiple \pidal\ 
decays, from a \piddal\ decay, or from photon conversions in the 
$(3.55\pm 0.17)\times 10^{-3}$ radiation length vacuum window assembly
\cite{vacwin}. 

The four-electron backgrounds fell into two categories. 1)~``Correctly 
paired'' four-track backgrounds, where all four electrons came from the 
same $\pi^0$:
these included i)~\piddal\ decays, ii)~\pidal\ where the photon from the 
Dalitz 
decay converted, iii)~$3\pi^0 \rightarrow 6\gamma$
where two photons from the same $\pi^0$ converted. In
correctly paired four-track events, the reconstructed \mee\ was generally
below $m_{\pi^0}$, and \mpipiee\ was slightly below $m_{K_L}$.
2)~``Mispaired'' four-track backgrounds, where the four
electrons came from {\em different} $\pi^0$'s:  These included
i)~events where two $\pi^0$'s decayed to $e^+e^-\gamma$, 
ii)~\pidal\ events where a photon from a different $\pi^0$
converted, iii)~$3\pi^0 \rightarrow 6\gamma$ events where two
photons from different $\pi^0$ decays converted. In these cases, because
the two observed electrons did not come from the same $\pi^0$, the
\mee\ distribution was nearly flat. Because the four photons were not
from two \pigg\ decays, the $Z$ position and the photon four-momenta were 
misreconstructed, giving a flat \mpipiee\ distribution as well.

Requiring the pairing $\chi^2$ to be below 4.5 removed 88\% of the
mispaired four-track background at the cost of 10\% of the signal.
Tightening the total mass cut to $|\mpipiee - m_{K_L}| < 10$~MeV/$c^2$
removed a further 80\% of the mispaired four-track background with 
negligible signal loss. 

The correctly paired four-track background could not be reduced significantly 
with pairing or kinematic cuts. About 99\% of these events, as well as 98\% 
of the remaining mispaired four-track background and 8\% of signal, were 
removed 
by cutting events with evidence of extra in-time tracks in the second drift 
chamber. The last three cuts were applied to both the signal and 
normalization samples.

\begin{figure}
  \centerline{ \psfig{figure=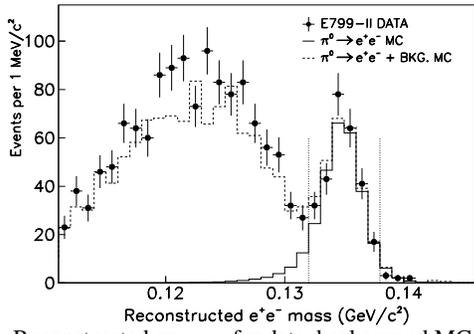,width=6.5cm}}
  \caption{Reconstructed \mee\ for data, background MC, and signal MC after 
           all cuts. The vertical
           dotted lines indicate the defined signal region 
           ($0.132<\mee<0.138$~GeV/$c^2$).
\label{mass1_prl}}
\end{figure}

After all cuts, the total background was dominated by high-mass 
Dalitz decays (18.1~$\pm$~4.7 events). Smaller backgrounds came from 
correctly paired 
(2.8~$\pm$~1.1 events) and mispaired (0.5~$\pm$~0.5 events) four-track 
modes. The errors on the four-track backgrounds are from MC
statistics; the error on the Dalitz background reflects MC statistics and
a 20\%  systematic uncertainty in the MC prediction of the misreconstructed 
\mee\ tail. An (18~$\pm$~5)\% discrepancy was seen between the data and the 
MC prediction in the level of the low-\mee\ Dalitz background between 
$0.110<\mee <0.125$~GeV/$c^2$ (Fig. \ref{mass1_prl}). Perfect agreement in 
this region was not expected, as these events typically had an extra 
low-energy photon near the cluster energy threshold. Although the Dalitz
decays which entered the signal sample had a much lower-energy photon and were
therefore less sensitive to the modeling of the threshold in the MC, we have 
treated this discrepancy conservatively as an additional systematic error 
on the background. The final background estimate was 
therefore 21.4~$\pm$~6.2 events. 

Radiative corrections to \pitoee\ had a significant effect on the acceptance.
Internal bremsstrahlung can produce a $e^+e^-\gamma$ final 
state with $\mee < m_{\pi^0}$, indistinguishable from the tree-level Dalitz 
decay. Following the convention of Ref. \cite{ksmcfprl}, we imposed a 
cutoff $\kwx > 0.95$. We thus quote the branching ratio for this range only, 
after subtracting the small contribution from the Dalitz diagram. In 
this kinematic region, interference between the two processes is negligible
\cite{bergstrom}. 

The normalization data set contained 650~264 events, with negligible 
backgrounds. The acceptance for Dalitz decays with $\mee > 65$~MeV/$c^2$
was 1.03\%, for kaons which decayed between 90
and 160~m from the target and had momentum between 20 and 200~GeV. The 
acceptance for 
the signal mode was 2.52\%. %Scaling by the ratio of the acceptances, 
%and the Dalitz decay branching ratio, the single-event sensitivity for 
%\pitoee, $\kwx>0.95$ was $2.40\times 10^{-10}$.

In the data, 275 \pitoee\ candidate events were observed (Fig. 
\ref{mass1_prl}). Subtracting the estimated background yielded the total 
sample size of 253.6~$\pm$~16.6 events (the error is statistical only). 

\begin{figure}
  \centerline{ \psfig{figure=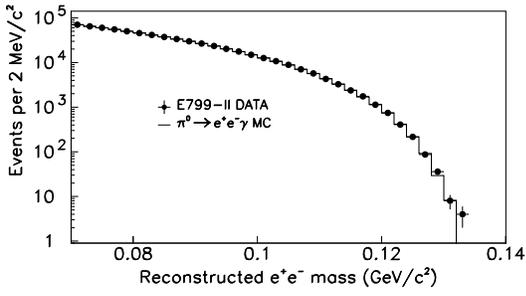,width=7.5cm}}
  \caption{Reconstructed \mee\ for normalization Dalitz decays.
\label{dalitzx_prl}}
\end{figure}

The largest source of systematic error was
%Significant systematic errors came from parameters of the Dalitz decays 
%used in the normalization, primarily 
the 2.7\% uncertainty in the \pidal\ 
branching ratio $(1.198\pm 0.032)\times 10^{-2}$ \cite{pdg}. In addition,
the \mee\ cutoff in the 
normalization Dalitz decays introduced a dependence of the acceptance
on the Dalitz decay form factor. The MC used the form factor slope of
0.033$~\pm~$0.003 measured by the CELLO collaboration \cite{cello}, 
which gives the
result that the $\mee > 65$~MeV/$c^2$ region contains 3.19\% of all
Dalitz decays. The CELLO measurement used the reaction 
$e^+e^- \rightarrow \pi^0e^+e^-$ in a region of spacelike momentum transfer,
extrapolating the slope to the kinematic region of the Dalitz decay 
assuming vector meson dominance. The most recent direct measurement from 
the Dalitz decay is consistent but less precise \cite{meijerdrees}.
Using the CELLO form factor, the observed \mee\ distribution 
(Fig. \ref{dalitzx_prl}) was consistent with the MC. The statistical
precision of our fit was 0.007, which we have taken to be the uncertainty in
the form factor; this translates into a 0.5\% systematic error on our
measurement of \pitoee .

The Dalitz decay branching ratio and the background uncertainty
dominated the systematic error. Smaller acceptance uncertainties included
a 1.0\% uncertainty in the efficiency of the pairing 
$\chi^2$ cut and a 1.2\% systematic error assigned to the efficiency 
of the \meeg\ cut in the normalization sample. These errors were determined
from resolution studies using fully-reconstructed Dalitz decays.
Adding all the systematic errors in quadrature, we obtained a total 
systematic error of 4.0\%. Our result for the branching ratio is
$ {\rm BR}\left (\pitoee , \kwx >0.95 \right) =
           (6.09 \pm 0.40 \pm 0.24) \times 10^{-8}, $
where the first error is statistical and the second systematic.

For comparison with the unitarity bound and with theoretical models which 
neglect final-state radiation, we can invert the radiative corrections and 
extrapolate our result to the
``lowest-order'' rate (what the branching ratio would be in the absence
of final-state radiation). This yields
$
\frac{\Gamma^{\rm lowest~order}_{e^+e^-}}{\Gamma_{\rm all}} =
   \left(7.04 \pm 0.46{\rm (stat)} \pm 0.28{\rm (syst)}\right)\times 10^{-8}
$, which is over four standard deviations above the unitarity bound.
This result, which is in agreement with recent Standard Model predictions,
represents the first statistically signficant observation of an excess
above unitarity.

E799-II expects to accumulate 2 to 4 times more data in a 1999 run, which
will allow a further refinement of this measurement. 
This result should provide constraints for predictions
of related decay modes such as $\eta \rightarrow \mu^+\mu^-$ and 
$K_L \rightarrow \mu^+\mu^-$ \cite{ametller,savage,pich,valencia}, and 
we hope that future experiments will be able to test these predictions.

This work was supported by U.S. D.O.E., N.S.F., and the Japan Ministry
of Education and Science.

%We gratefully acknowledge the support and effort of the Fermilab
%staff and the technical staffs of the participating institutions for
%their vital contributions.  This work was supported in part by the U.S. 
%Department of Energy, The National Science Foundation and The Ministry of
%Education and Science of Japan. 
%In addition, A.R.B., E.B. and S.V.S. 
%acknowledge support from the NYI program of the NSF; A.R.B. and E.B. from 
%the Alfred P. Sloan Foundation; E.B. from the OJI program of the DOE; 
%K.H., T.N. and M.S. from the Japan Society for the Promotion of
%Science.  P.S.S. acknowledges receipt of a Grainger Fellowship.

% now the references. delete or change fake bibitem. delete next three
%   lines and directly read in your .bbl file if you use bibtex.

\end{document}